\documentclass[manuscript]{aastex}
\usepackage{color}

\slugcomment{Submitted to Astrophysical Journal in 8th of December 2014}

\shorttitle{Transit polarization in exoplanetary systems}
\shortauthors{Kostogryz, Yakobchuk \& Berdyugina}

\begin{document}

\title{Polarization in exoplanetary systems caused by transits, grazing transits, and starspots}

\author{N.M. Kostogryz\altaffilmark{1,2}, T.M.Yakobchuk\altaffilmark{1,2}, S.V.Berdyugina\altaffilmark{1} }
\affil{$^1$Kiepenheuer-Institute f\"{u}r Sonnenphysik}
 \affil{Sch\"{o}neckstr.6, D-79104, Freiburg, Germany }
 \email {kostogryz@kis.uni-freiburg.de}
 \email {sveta@kis.uni-freiburg.de}

  \affil{$^2$Main Astronomical Observatory of NAS of Ukraine}
 \affil {Zabolotnoho str. 27, 03680, Kyiv, Ukraine}

\begin{abstract}

We present results of numerical simulations of flux and linear polarization variations
in transiting exoplanetary systems, caused by the host star disk symmetry breaking. 
We consider different configurations of planetary transits depending on orbital 
parameters. Starspot contribution to the polarized signal is also estimated. 
Applying the method to known systems and simulating observational
conditions, a number of targets is selected where transit polarization effects could be 
detected. We investigate several principal benefits of the transit polarimetry, particularly,
for determining orbital spatial orientation and distinguishing between grazing and near-grazing
planets. 
Simulations show that polarization parameters are also sensitive to starspots, and
they can be used to determine spot positions and sizes.

\end{abstract}


\section{Introduction}

The discovery of the first extrasolar planet has led to the rapid development of many new methods
for their detection and characterization. The best-characterized planets so far
are those which were observed and analyzed by using multiple methods. Thanks to the recent high accuracy
instruments, polarimetry has become a promising technique for characterizing
exoplanetary systems that can yield information inaccessible to other methods. Particularly,
\citet{berd08, fluri10, berd11}
showed that polarization variability which occurs due to scattering in the planetary atmosphere
over the orbital period is
capable to reveal the orbital period of the planet, inclination, eccentricity, orientation of the
orbit in space as well as the nature of scattering particles in the planetary atmosphere.

Another interesting polarimetric effect is expected in exoplanetary systems when transits take place.
By breaking the symmetry of the intensity integrated over the stellar disk, a transiting planet 
results in linear polarization of a partially eclipsed star. Such an effect 
was first detected in the eclipsing binary Algol \citep{kemp83}.
First estimates of the effect for transiting planets have been made only recently. \citet{carciofi05}
numerically simulated the occultation polarization in exoplanetary systems of G dwarf stars,
with planet sizes ranging from that of 1 to 2 times the size of Jupiter. They also explored
the cases of K-M-T dwarfs with a simplified limb polarization approximation and suggested that in
favorable situations ground-based polarimetry should be able to detect Earth-like planets orbiting
these stars. 

\citet{kostogryz11} using the same Monte Carlo technique, 
simulated the transit polarization for several objects with high planet-to-star radii ratio. 
For the TrES-3, WASP-4 and WASP-25 systems with G spectral type host stars the modelled 
solar limb polarization \citep{trujillo09} was
chosen, while for one of the brightest transit exoplanetary system HD~189733 (K spectral type) pure
scattering atmosphere approximation \citep{chandr60} was 
applied. Obviously, the latter approximation is too rough to predict the observational values of
transit polarization. Thus, \citet{frantseva12}, assuming solar limb polarization for HD~189733,
obtained the polarization that is different by almost two orders of magnitude as compared to
\citet{kostogryz11}, showing the importance of knowing the intrinsic stellar polarization.

The center-to-limb variation of the linear polarization across the stellar disk originates
from scattering opacity contributing to the total opacity in the atmosphere.
By solving the radiative transfer equation for polarized light, 
\citet{kostogryz14} calculated center-to-limb variations of the intensity and the linear
polarization for stars of different spectral types accounting for various opacity sources.
It was shown that low-gravity cool stars, while having weaker
limb darkening, should exhibit significantly larger linear polarization in their continuum
spectra as compared to solar-type stars. At the same time, for stars with the same 
effective temperature, the limb polarization is larger for the lower gravity stars, i.e. giants.

To detect a tiny polarization arizing from stellar symmetry breaking effect caused by a transiting exoplanet, 
a very sensitive polarimeter should be used. The most accurate polarimetric observations
were made for the Sun at a sensitivity level of parts in ten millions $10^{-7}$ by \citet{kemp87}. Up to now, 
there are no more polarimeters which can get that high sensitivity level. Nevertheless, there are several
of them coming close. The Zurich Imaging Polatimeter (ZIMPOL) has achieved the $10^{-6}$ relative 
sensitivity in solar spectropolarimetry \citep{stenflo00} but its absolute accuracy is unknown, ZIMPOL is also one of the 
components for the SPHERE \citep{beuzit06} that can reach a polarimetric sensitivity about $10^{-5}$ \citep{thalmann08}.
PlanetPol can reach $10^{-6}$ on bright target but its 
systematics is larger than the internal errors by at least factor 1.8 \citep{hough06, lucas09, bailey10}. 
The broad-band polarimeter DiPOL-2 is able to measure a polarized signal with a high precision
of $10^{-5}$ which appears photon limited on small telescopes. DiPOL-2 has the same statistic 
and systematic errors \citep{piirola14} which was demonstrated by reproducing the measurements for Algol by 
\citet{kemp83} using a 60-cm telescope.

In this paper, we present our results of modeling the transit polarimetric effect for
known transiting exoplanets which orbit spotted stars using the modelled center-to-limb variations of the
intensity and linear polarization  by \citet{kostogryz14}. In Section 2 we describe our semi-analytical method
for modeling flux and linear polarization curves during planetary transits.
In Section 3 the results of our calculations for 88 known transiting exoplanets are presented. Several 
exoplanetary systems that have grazing and near-grazing transiting planets are studied in Section 4. 
As a spot on the star can produce 
the polarimetric signal due to the same symmetry breaking effect, we investigated it in the Section 5.
Finally, we present a summary of this paper in Section 6.

\section{Method of transit polarization calculations}

\subsection{Basic equations.}
For a given set of parameters describing configuration of the planetary system,
the residual normalized stellar flux when the star is blocked by the planet can be expressed as,

\begin{equation}
F(\bar {p}_0, \phi_0)~=~1-\frac{1}{\pi}~\int_{0}^{2\pi} d\phi'~ \int_{0}^{R_p/R_\star}\bar{p}'~f(\bar{p}', \phi')~d\bar{p}'
\label{Eq:flux}
\end{equation} 

\noindent where $R_p$ and $R_\star$ are the planetary and stellar radii, respectively, $\bar{p}'$ and 
$\phi'$ are polar coordinates of a stellar disk, $\bar{p}_0$ and $\phi_0$
are the polar coordinates of the center of the planet, $f(\bar{p}', \phi')$ is the stellar limb darkening.

The normalized Stokes parameters are
 
\begin{mathletters} 
 \begin{eqnarray}
 q(\bar {p}_0, \phi_0)~=~\frac{1}{\pi}~\int_{0}^{2\pi} d\phi'~ \int_{0}^{R_p/R_\star}\bar{p}' \times  \nonumber \\ 
 					\times~f(\bar{p}', \phi')~P(\bar{p}', \phi')~\cos(2\phi)~d\bar{p}'  \\
 u(\bar {p}_0, \phi_0)~=~-\frac{1}{\pi}~\int_{0}^{2\pi} d\phi'~ \int_{0}^{R_p/R_\star}\bar{p}'\times~ \nonumber \\
 					\times~f(\bar{p}', \phi')~P(\bar{p}', \phi')~\sin(2\phi)~d\bar{p}' \label{Eq:pol}
\end{eqnarray} 
\end{mathletters}

where $P(\bar{p}', \phi')$ is the center-to-limb variation of the linear polarization of the star. Here q (North-South)
and u (45 $\deg$ counterclockwise to q) are defined as positive. 

As seen from Eq. \ref{Eq:flux} and Eq. 2(a,b), to calculate the flux and two normalized Stokes 
parameters q and u during the transit, the center-to-limb variations of the intensity $f(\bar{p}', \phi')$
and the linear polarization $P(\bar{p}', \phi')$ should be known.

 \subsection{Stellar limb darkening and limb polarization.}
 
 We calculate the center-to-limb variation of intensity (CLVI) and linear polarization (CLVP) 
in the continuum spectra for the grid of Phoenix models \citep{hauschildt99} within the range
 of effective temperatures from 3500K to 6000K and for gravity $\log g = 4.0 - 5.0$.  
 Assuming no magnetic field present, we solved the radiative transfer 
 equations for polarized light iteratively considering plane-parallel model atmosphere and various opacities. 
 The method of radiative transfer equations and opacity contributions calculation are described in \citet{kostogryz14}. 

\begin{figure}
\epsscale{.90}
\plotone{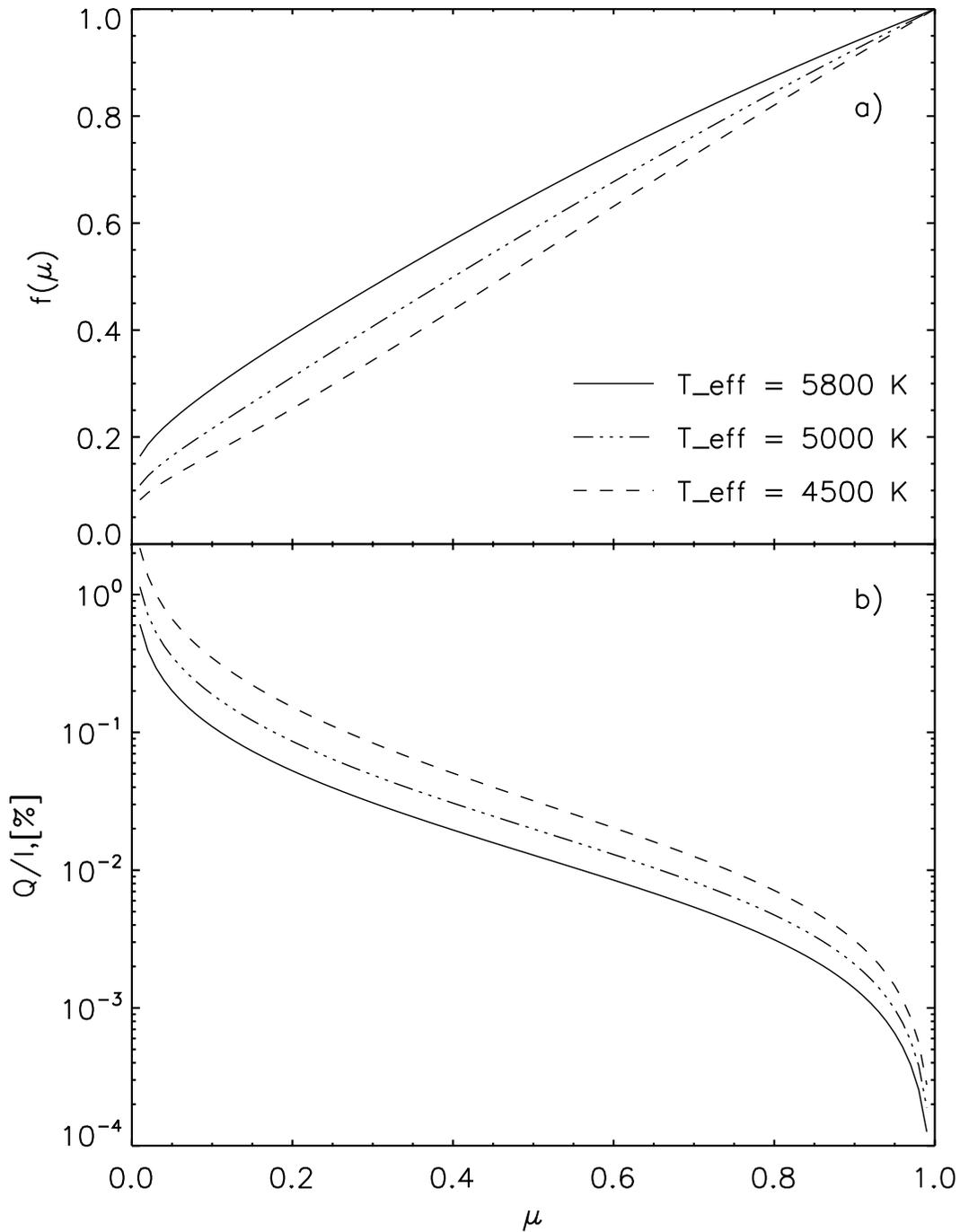}
\caption{Center-to-limb variation of the intensity (a) and the linear polarization (b) in the continuum at $4500\rm{\AA}$ for 
$\log g = 4.5$. Different types of lines correspond to different temperatures labeled in (a) panel. 
The data for these plots were calculated in \citet{kostogryz14}.
\label{fig1}}
\end{figure}

\placefigure{fig1}
 
In Fig. \ref{fig1} we present some examples of intensity (Fig. \ref{fig1},a) and polarization variation (Fig. \ref{fig1},b)
depending on the position on the stellar disk $\mu = \cos \phi'$ (CLVI and CLVP)
for models with effective temperatures of 5800K, 5000K, 4500K,  $\log g = 4.5$ 
and $\lambda = 4500 \rm{\AA}$. The $\log g$ was chosen to be 4.5 since most of the known
transiting exoplanetary hosts have $\log g $ in the range between 4.3 to 4.7. As shown 
in \citet{kostogryz14}, the CLVP and CLVI strongly depend on the value of $\log g$. Using 
the look-up tables we interpolate them to get the 
CLVI and the CLVP for the host star of particular effective temperature and surface gravity.

\subsection{Implementation}  
 
In the previous studies \citep{kostogryz11, frantseva12} the method of \citet{carciofi05}
to model the polarization effects in transiting exoplanet systems was employed. While being
simple, this Monte Carlo based approach is flexible and can be easily modified to
account for different cases such as spots or stellar oblateness. However,
since considered polarimetric effects are often small one needs to use
a huge number of photon packets emitted from randomly chosen locations on
a host star in order to obtain smooth Stokes parameter variation curves in the
output. This makes the method time-consuming and inapplicable for solving
the inverse problem.

Current methods for transit light curve modeling \citep[e.g.,][]{eastman13} are heavily
based on the analytical models from \citet{mandel02}. They are much faster
than one- or two-dimensional numerical integration methods. However, analytical
models require analytical functions for limb-darkening and, in our case,
for polarization as well.
While we tried different functions to fit calculated limb-darkening and polarization
variations, none were successful enough for our task, resulting
in discernible artifacts, especially, on the Stokes parameters variation curves.
Thus, we interpolated limb-darkening and polarization from the
look-up tables that were calculated with steps by $\mu$ equal or better than
0.01 \citep{kostogryz14} and numerically integrated over the stellar surface.

\begin{figure}
\includegraphics[width=.5\linewidth,angle=-90]{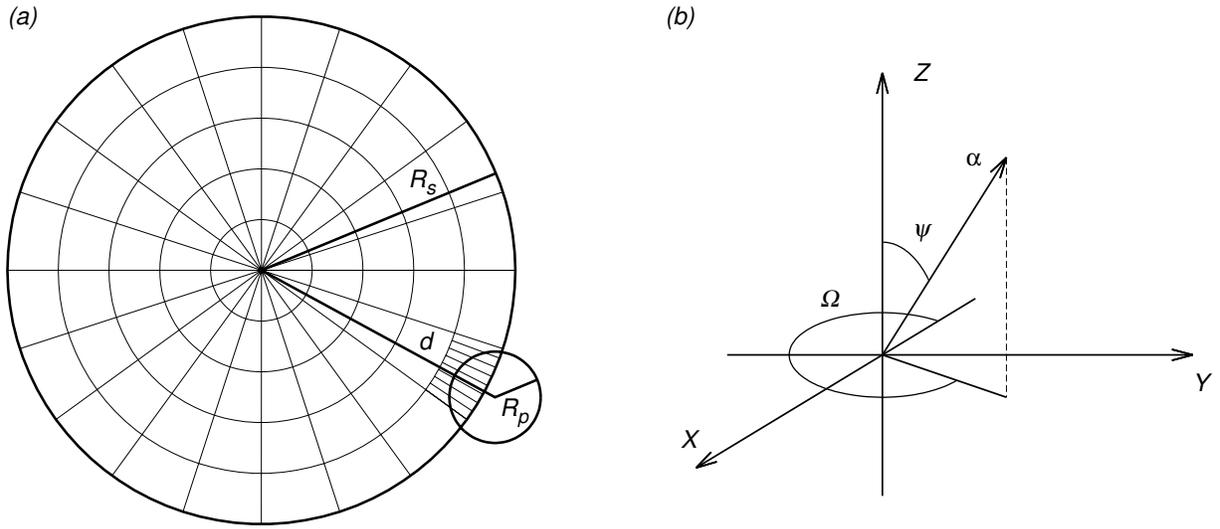}
\caption{a) Example polar grid used in the transit simulations.
The pixel behind the planet demonstrates subdivision of partially 
covered pixels into sectors for more accurate area calculations.
 b) A coordinate system in which \textbf{$\alpha$} is the
 host star rotational axis orientation, and the X-Y-plane is
the orbital plane that is described in \citet{fabrycky09}. Note that here 
$\Omega$ is the azimuthal angle of rotational axis of the star 
and not the ascending node longitude discussed in the text.
 \label{fig2}}
\end{figure}

\placefigure{fig2}

In our method we use polar grid pixelization as shown in Figure \ref{fig2}a.
The polar coordinate system corresponds to the stellar disk sky projection and is centered
on the disk center. We have tested different grids, but ended up with a
uniform grid in both radial and angular coordinates. All distances and sizes are
normalized to the host star's radius. After initializing the grid we pre-calculate
for each surface pixel the area, the flux and Stokes polarization parameters from
our limb-darkening and polarization tables using 3D linear interpolation in
effective temperature, gravity and wavelength. The basic idea of modeling is to
exclude from the integration those pixels that are covered by the passing
opaque planet. The finite size pixelization requires a special treatment
for partially covered pixels. Since enlargement of the grid dimensions worsen 
the computational speed, for such pixels we employ subdivision into smaller sectors by 
polar angle (see Fig. \ref{fig2}a). We assign the same pre-calculated flux and the
polarization for the partially covered pixels scaled by the covered-to-open area ratio.
The calculation of the covered subpixel area is made with trapezoids.

After a number of tests we have chosen 200x200 pixel grid with 100 subpixel division.
Our comparison with the \citet{carciofi05} method shows a significant
gain in performance without loss of accuracy. We think that with an appropriate
minimization algorithm our method could be used for solving the inverse problem
to obtain different system parameters using polarization data in addition to
lightcurves. Each transit flux and polarization curves were calculated for 200 
points across the event to allow the maximum polarization to be well seen. However, to simulate more 
realistic observations, we averaged the Stokes parameters assuming that we can 
get only 10 for each q and u measurements per transit, because
the exposure time cannot be very small, as we need to integrate longer
to achieve a required sensitivity.

We have realized both circular and elliptical orbital motion in the program. The later
involves Kepler's equation solution as described in \citet{fukushima97}.
Since the transit usually lasts for only a few hours, which is short compared to
the orbital period, orbital ellipticity does not change the results significantly.
However, the orientation of the orbit measured by longitude of the ascending node $\Omega$
expectedly alters the curves for q and u Stokes parameters (see below). As 
for most exoplanetary systems the $\Omega$ is not known, we assume it equal to $90~\deg$. 

Along with the planet transit we have accounted in our code for spots and
binary system modeling (with possibility to add spots on both stars). The core
integration algorithm is the same as described above. Spots are considered under
the following simple assumptions: they have circular boundary with no penumbra,
show no temporal size and position changes, and there is no differential rotation
of the star. According to the assigned temperature for each spot a separate grid
of flux and polarization pixel values is pre-calculated for spots. For partially covered 
pixels we account for both covering by planet and spot. For the
host star we set the axis of rotation orientation following the Rossiter-McLaughlin
effect conventions shown in Figure \ref{fig2}b \citep[for details, see e.g.][]{fabrycky09}.
In the sections below we analyze the modeling results with our method of
different test cases and real exoplanet systems.


\section{Planetary transits}

The main advantage of polarimetry is that we have two parameters (Stokes Q and U)
that are varying and can change their sign during the transit, providing more
information than photometry, which measures only one parameter -- the relative flux.
Measuring variations of normalized Stokes $Q/I = q$ and $U/I = u$,
one can also get the longitude of the ascending node $\Omega$ in addition to the 
inclination angle, that can be determined from photometric observations. 

In addition to the planetary parameters,
polarimetry yields information about the host star as well. Particularly,
it can give the center-to-limb variation of the polarization, which could be used
for testing the stellar models.

There are more than 600 transiting exoplanetary systems with various parameters known to date. 
Based on those parameters that contribute to the transit polarization,
we have formulated a few criteria for larger polarization and selected promising 
exoplanets mostly using the catalog from the Extrasolar Planets Encyclopaedia (\textit{http://exoplanet.eu})
and for some cases that did not have enough information we use additional Exoplanets 
Orbit Database (\textit{http://exoplanets.org}).
In cooler stellar atmosphere a contribution
to the opacity from Rayleigh scattering increases that gives rise to polarization at the 
stellar limb \citep{harrington69, magalhaes86,kostogryz14}.

Here we constrained the effective temperature of the host stars to be in the range
between 3500K and 6000K. Another important parameter that leads to larger polarization 
during transit is the planet-to-star radii ratio, which was
set to be larger than 0.1. According to these criteria, we obtained a sample of 88 transiting
exoplanets presented in Table \ref{tbl-1}. For each system, we list the planet-to-star radii ratio
($R_p/R_\star$), the effective temperature $T_{\rm eff}$, the surface gravity $\log g$,
the orbit inclination angle $i$, the apparent magnitude in V filter, and the maximum continuum
polarization degree $P_{\rm max}$ for light curves with 10 points at wavelengths 
$\lambda = 4000\rm{\AA}, 4500\rm{\AA}$  and $5000\rm{\AA}$ 
predicted by our models. We list $P_{\rm max}$ instead of $q_{\rm max}$ and $u_{\rm max}$
because the positional angle of the orbit $\Omega$ is not known in most cases. It can 
be determined from polarization measurements taken during the entire orbit \citep{fluri10}.
Since the intrinsic polarization of stars is the largest
at the limb, the maximum polarization during the transit occurs when the
planet eclipses the edge of the stellar disk. The shape of the transit curves depend on the orbital parameters.

\placetable{tbl-1}
\begin{deluxetable}{lrrrrrrrr}
\tablecolumns{9}
\tablewidth{0pt}
\tablecaption{Maximum polarization degree during planetary transit of different systems
with different stellar parameters at $\lambda~ 4000 \rm{\AA}$, $4500 \rm{\AA}$ and $5000 \rm{\AA}$. 
Almost all systems from our list have only one transiting planet, except of HAT-P-44 for 
which we have chosen HAT-P-44b planet.
The averaged number of measurements per transit is equal to 10. \label{tbl-1}}
\tablehead{
\colhead{Star} & \colhead{$R_p/R_\star$} & \colhead{$T_{eff}$} & \colhead{$\log g$} & \colhead{$i, \deg$} & \colhead{$m_v$} &\multicolumn{3}{c}{$P_{max}, 10^{-6}$} \\
 \cline{7-9} \\
\colhead{} & \colhead{}   & \colhead{}    & \colhead{} &  \colhead{} &
\colhead{}  & \colhead{$\lambda = 4000\rm{\AA}$}    & \colhead{$\lambda = 4500\rm{\AA}$} & \colhead{$\lambda = 5000\rm{\AA}$}}
\startdata

CoRoT-10 & 0.12606 & 5075 & 4.591 & 88.55 & 15.2 & 5.33 & 2.74 & 1.50 \\
CoRoT-12 & 0.13247 & 5675 & 4.374 & 85.45 & 15.5 & 4.70 & 2.42 & 1.33 \\
CoRoT-16 & 0.10094 & 5650 & 4.327 & 85.10 & 15.6 & 3.85 & 1.98 & 1.09 \\
CoRoT-18 & 0.13450 & 5440 & 4.415 & 86.50 & 15.0 & 5.17 & 2.66 & 1.46 \\
CoRoT-2 & 0.16675 & 5575 & 4.513 & 87.84 & 12.6 & 5.97 & 3.07 & 1.68 \\
CoRoT-9 & 0.10267 & 5625 & 4.486 & 89.90 & 13.7 & 2.68 & 1.37 & 0.75 \\
HAT-P-1 & 0.11535 & 5980 & 4.359 & 86.28 & 10.3 & 4.56 & 2.35 & 1.29 \\
HAT-P-15 & 0.10191 & 5568 & 4.376 & 89.10 & 12.2 & 3.24 & 1.66 & 0.91 \\
HAT-P-17 & 0.12389 & 5246 & 4.525 & 89.20 & 10.5 & 4.39 & 2.25 & 1.23 \\
HAT-P-18 & 0.13574 & 4803 & 4.575 & 88.80 & 12.8 & 6.60 & 3.40 & 1.86 \\
HAT-P-19 & 0.14173 & 4990 & 4.535 & 88.20 & 12.9 & 6.27 & 3.23 & 1.77 \\
HAT-P-20 & 0.12826 & 4595 & 4.633 & 86.80 & 11.3 & 6.53 & 3.36 & 1.83 \\
HAT-P-22 & 0.10662 & 5302 & 4.365 & 86.90 & 9.7 & 4.04 & 2.08 & 1.14 \\
HAT-P-23 & 0.10888 & 5924 & 4.269 & 85.10 & 11.9 & 3.93 & 2.03 & 1.12 \\
HAT-P-25 & 0.12740 & 5500 & 4.478 & 87.60 & 13.2 & 4.32 & 2.22 & 1.22 \\
HAT-P-27/WASP-40 & 0.12062 & 5300 & 4.506 & 84.98 & 12.2 & 4.73 & 2.43 & 1.33 \\
HAT-P-28 & 0.11281 & 5680 & 4.363 & 88.00 & 13.0 & 3.74 & 1.92 & 1.05 \\
HAT-P-3 & 0.10627 & 5224 & 4.594 & 87.07 & 11.6 & 3.37 & 1.73 & 0.95 \\
HAT-P-32 & 0.15078 & 6001 & 4.223 & 88.70 & 11.3 & 9.06 & 4.70 & 2.60 \\
HAT-P-36 & 0.11841 & 5580 & 4.367 & 86.00 & 12.3 & 4.21 & 2.16 & 1.18 \\
HAT-P-37 & 0.13791 & 5500 & 4.519 & 86.90 & 13.2 & 4.57 & 2.35 & 1.29 \\
HAT-P-43 & 0.11931 & 5645 & 4.371 & 88.70 & 13.4 & 4.07 & 2.09 & 1.14 \\
HAT-P-44 & 0.13423 & 5295 & 4.428 & 89.00 & 13.4 & 5.36 & 2.75 & 1.50 \\
HAT-P-5 & 0.11305 & 5960 & 4.391 & 86.75 & 12.0 & 3.84 & 1.98 & 1.10 \\
HAT-P-54 & 0.15708 & 4390 & 4.666 & 87.04 & 13.5 & 11.97 & 6.10 & 3.30 \\
HATS-1 & 0.12878 & 5870 & 4.398 & 85.60 & 12.1 & 3.96 & 2.05 & 1.13 \\
HATS-2 & 0.13354 & 5227 & 4.476 & 87.20 & 13.6 & 5.24 & 2.69 & 1.47 \\
HATS-4 & 0.11321 & 5403 & 4.505 & 88.50 & 13.5 & 3.44 & 1.77 & 0.97 \\
HATS-5 & 0.10750 & 5304 & 4.528 & 89.30 & 12.6 & 3.24 & 1.66 & 0.90 \\
HATS-6 & 0.17976 & 3724 & 4.684 & 88.21 & 15.2 & 8.35 & 3.95 & 2.02 \\
HD-189733 & 0.14827 & 4980 & 4.547 & 85.76 & 7.7 & 7.24 & 3.72 & 2.03 \\
KIC-6372194 & 0.10130 & 5233 & 4.590 & 89.69 & 16.3 & 3.03 & 1.55 & 0.85 \\
Kepler-12 & 0.11734 & 5947 & 4.161 & 88.76 & 13.8 & 5.49 & 2.84 & 1.57 \\
Kepler-16AB & 0.11906 & 4450 & 4.741 & 90.03 & 12.0 & 6.06 & 3.10 & 1.68 \\
Kepler-17 & 0.12829 & 5781 & 4.459 & 87.20 & 13.8 & 3.99 & 2.05 & 1.12 \\
Kepler-30 & 0.11888 & 5498 & 4.477 & 89.92 & 15.5 & 3.84 & 1.97 & 1.08 \\
Kepler-412 & 0.10570 & 5750 & 4.285 & 80.89 & 13.7 & 4.08 & 2.10 & 1.15 \\
Kepler-45 & 0.17920 & 3820 & 4.727 & 87.00 & 16.9 & 10.00 & 4.67 & 2.36 \\
Kepler-71 & 0.13151 & 5591 & 4.579 & 89.95 & 15.4 & 3.87 & 1.99 & 1.09 \\
Kepler-75 & 0.12017 & 5330 & 4.493 & 89.10 & 15.0 & 4.02 & 2.06 & 1.12 \\
OGLE-TR-111 & 0.13306 & 5070 & 4.512 & 88.10 & 17.0 & 5.53 & 2.84 & 1.56 \\
OGLE-TR-113 & 0.14897 & 4752 & 4.562 & 89.40 & 16.1 & 8.01 & 4.13 & 2.26 \\
OGLE-TR-182 & 0.13239 & 5924 & 4.380 & 85.70 & 16.8 & 5.13 & 2.64 & 1.45 \\
Qatar-1 & 0.14521 & 4861 & 4.536 & 84.52 & 12.8 & 6.82 & 3.51 & 1.91 \\
Qatar-2 & 0.16473 & 4645 & 4.600 & 88.30 & 13.3 & 9.93 & 5.13 & 2.80 \\
TrES-1 & 0.13575 & 5230 & 4.567 & 88.40 & 11.8 & 4.88 & 2.50 & 1.37 \\
TrES-2 & 0.12207 & 5850 & 4.428 & 83.62 & 11.4 & 4.23 & 2.18 & 1.20 \\
TrES-3 & 0.16480 & 5720 & 4.582 & 82.15 & 12.4 & 6.09 & 3.12 & 1.71 \\
TrES-5 & 0.14333 & 5171 & 4.513 & 84.53 & 13.7 & 5.40 & 2.76 & 1.50 \\
WASP-10 & 0.14161 & 4675 & 4.501 & 86.80 & 12.7 & 7.76 & 4.01 & 2.19 \\
WASP-104 & 0.12552 & 5450 & 4.509 & 83.63 & 11.1 & 4.58 & 2.35 & 1.29 \\
WASP-11/HAT-P-10 & 0.13245 & 4980 & 4.534 & 88.50 & 11.9 & 5.70 & 2.93 & 1.60 \\
WASP-16 & 0.10940 & 5550 & 4.495 & 85.22 & 11.3 & 3.67 & 1.88 & 1.03 \\
WASP-19 & 0.14265 & 5500 & 4.390 & 80.80 & 12.6 & 5.79 & 2.96 & 1.61 \\
WASP-2 & 0.13283 & 5150 & 4.519 & 84.73 & 11.9 & 5.82 & 2.99 & 1.63 \\
WASP-20 & 0.10761 & 5950 & 4.230 & 85.57 & 10.7 & 4.49 & 2.32 & 1.28 \\
WASP-21 & 0.11720 & 5800 & 4.391 & 87.90 & 11.6 & 3.84 & 1.98 & 1.09 \\
WASP-23 & 0.12911 & 5150 & 4.562 & 88.39 & 12.7 & 4.78 & 2.46 & 1.34 \\
WASP-25 & 0.13617 & 5750 & 4.482 & 87.70 & 11.9 & 4.37 & 2.25 & 1.23 \\
WASP-29 & 0.10210 & 4800 & 4.499 & 88.80 & 11.3 & 4.22 & 2.17 & 1.19 \\
WASP-34 & 0.13468 & 5700 & 4.504 & 85.20 & 10.3 & 2.64 & 1.37 & 0.76 \\
WASP-36 & 0.13816 & 5881 & 4.497 & 83.65 & 12.7 & 4.27 & 2.19 & 1.20 \\
WASP-37 & 0.12190 & 5800 & 4.386 & 88.78 & 12.7 & 3.94 & 2.03 & 1.11 \\
WASP-39 & 0.14569 & 5400 & 4.502 & 87.83 & 12.1 & 5.23 & 2.69 & 1.48 \\
WASP-4 & 0.12454 & 5500 & 4.284 & 89.35 & 12.5 & 5.04 & 2.59 & 1.41 \\
WASP-41 & 0.12300 & 5450 & 4.406 & 87.30 & 11.6 & 4.54 & 2.33 & 1.28 \\
WASP-43 & 0.15947 & 4520 & 4.644 & 82.60 & 12.4 & 11.19 & 5.76 & 3.13 \\
WASP-44 & 0.12626 & 5410 & 4.481 & 86.02 & 12.9 & 4.28 & 2.20 & 1.21 \\
WASP-45 & 0.12603 & 5140 & 4.445 & 84.47 & 12.0 & 6.04 & 3.11 & 1.70 \\
WASP-46 & 0.14667 & 5620 & 4.493 & 82.63 & 12.9 & 5.60 & 2.87 & 1.57 \\
WASP-47 & 0.10267 & 5400 & 4.351 & 89.20 & 11.9 & 3.42 & 1.75 & 0.96 \\
WASP-5 & 0.11091 & 5700 & 4.367 & 85.80 & 12.3 & 3.66 & 1.88 & 1.03 \\
WASP-50 & 0.13665 & 5400 & 4.508 & 84.74 & 11.6 & 4.95 & 2.54 & 1.38 \\
WASP-52 & 0.16505 & 5000 & 4.581 & 85.35 & 12.0 & 7.93 & 4.07 & 2.22 \\
WASP-55 & 0.12591 & 5900 & 4.391 & 89.20 & 11.8 & 3.98 & 2.05 & 1.13 \\
WASP-58 & 0.12022 & 5800 & 4.274 & 87.40 & 11.7 & 4.53 & 2.34 & 1.29 \\
WASP-6 & 0.14444 & 5450 & 4.506 & 88.47 & 12.4 & 5.13 & 2.63 & 1.44 \\
WASP-67 & 0.16521 & 5200 & 4.497 & 85.80 & 12.5 & 7.67 & 3.94 & 2.16 \\
WASP-77A & 0.13008 & 5500 & 4.478 & 89.40 & 11.3 & 4.32 & 2.22 & 1.21 \\
WASP-8 & 0.11183 & 5600 & 4.493 & 88.52 & 9.8 & 3.33 & 1.71 & 0.94 \\
WASP-80 & 0.15514 & 4145 & 4.602 & 89.92 & 11.7 & 11.67 & 5.74 & 3.02 \\
WASP-84 & 0.12930 & 5314 & 4.615 & 88.37 & 10.1 & 3.99 & 2.04 & 1.11 \\
WASP-95 & 0.10994 & 5630 & 4.376 & 88.40 & 10.1 & 3.51 & 1.80 & 0.99 \\
WASP-96 & 0.11733 & 5540 & 4.420 & 85.60 & 12.2 & 4.13 & 2.11 & 1.15 \\
WASP-97 & 0.10945 & 5640 & 4.436 & 88.00 & 10.6 & 3.26 & 1.67 & 0.91 \\
WASP-98 & 0.16134 & 5525 & 4.586 & 86.30 & 13.0 & 6.16 & 3.16 & 1.73 \\
WTS-2 & 0.17539 & 5000 & 4.588 & 83.43 & - & 8.87 & 4.56 & 2.48 \\
XO-2 & 0.10363 & 5340 & 4.460 & 88.70 & 11.2 & 3.19 & 1.63 & 0.89 \\
\enddata
\end{deluxetable}

\begin{figure}
\includegraphics[width=0.9\linewidth,angle=-90]{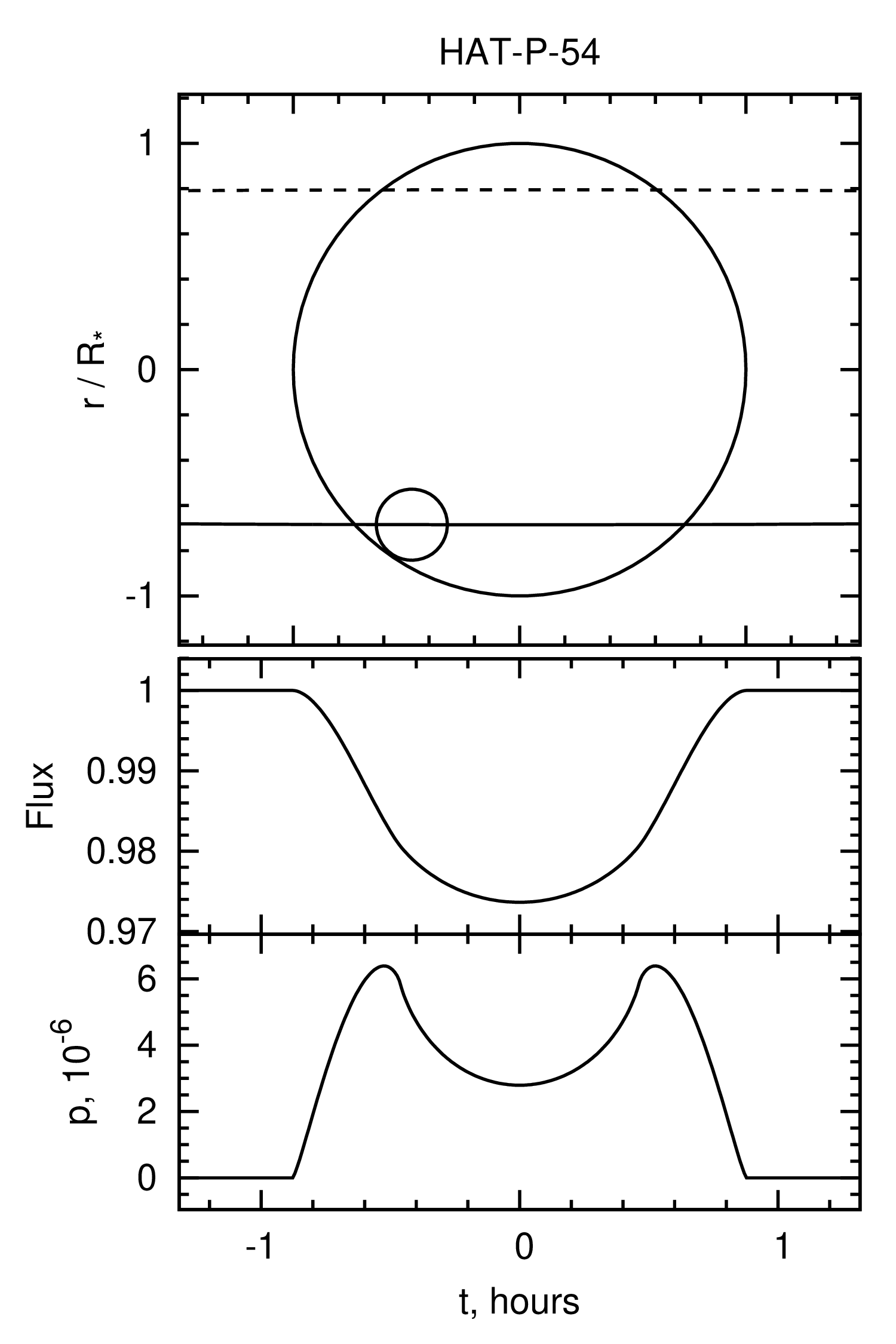}
\caption{ A configuration of the exoplanetary system, its flux and polarization degree 
assuming 200 measurements per transit at the wavelength $4500~\rm \AA$. 
The star and the planet are shown in the top left panel as large and small circles,
respectively. Parts of the planetary orbit are shown by horizontal lines. Solid 
line describes the planetary orbit in front of the star and dashed line shows the planetary orbit behind the star.
The planet is orbiting from left to right in front of the star.
Adopted positive q and u orientation
on the sky are shown in the top of the plot. Here the plots are only for the HAT-P-54 
system. All other systems from Table \ref{tbl-1} are shown in the online version of Fig. \ref{fig2.5}. 
\label{fig2.5}}
\end{figure}

We present fine-grid (200 points) flux and polarization degree for $\lambda = 4500\rm \AA$ 
during the planetary transit, together with the star-planet configuration, in Figure \ref{fig2.5}  
for all systems listed in Table \ref{tbl-1}. 
As Stokes parameters q and u depend on the longitude of ascending node
of the planetary orbit and direction of the planet motion around the star  
(see, for example, Fig. \ref{fig5.1} ) which are mostly unknown, we present 
here only the polarization degree which is independent on these parameters.
Note that maximum 
polarization degree for each planet will be slightly different than that in Table 1, as
in the table we present averaged polarization degree for 10 data points per transit, 
while in all plots we present 200 measurements per transit. 

\placefigure{fig2.5}

\begin{figure}
\epsscale{.75}
\plotone{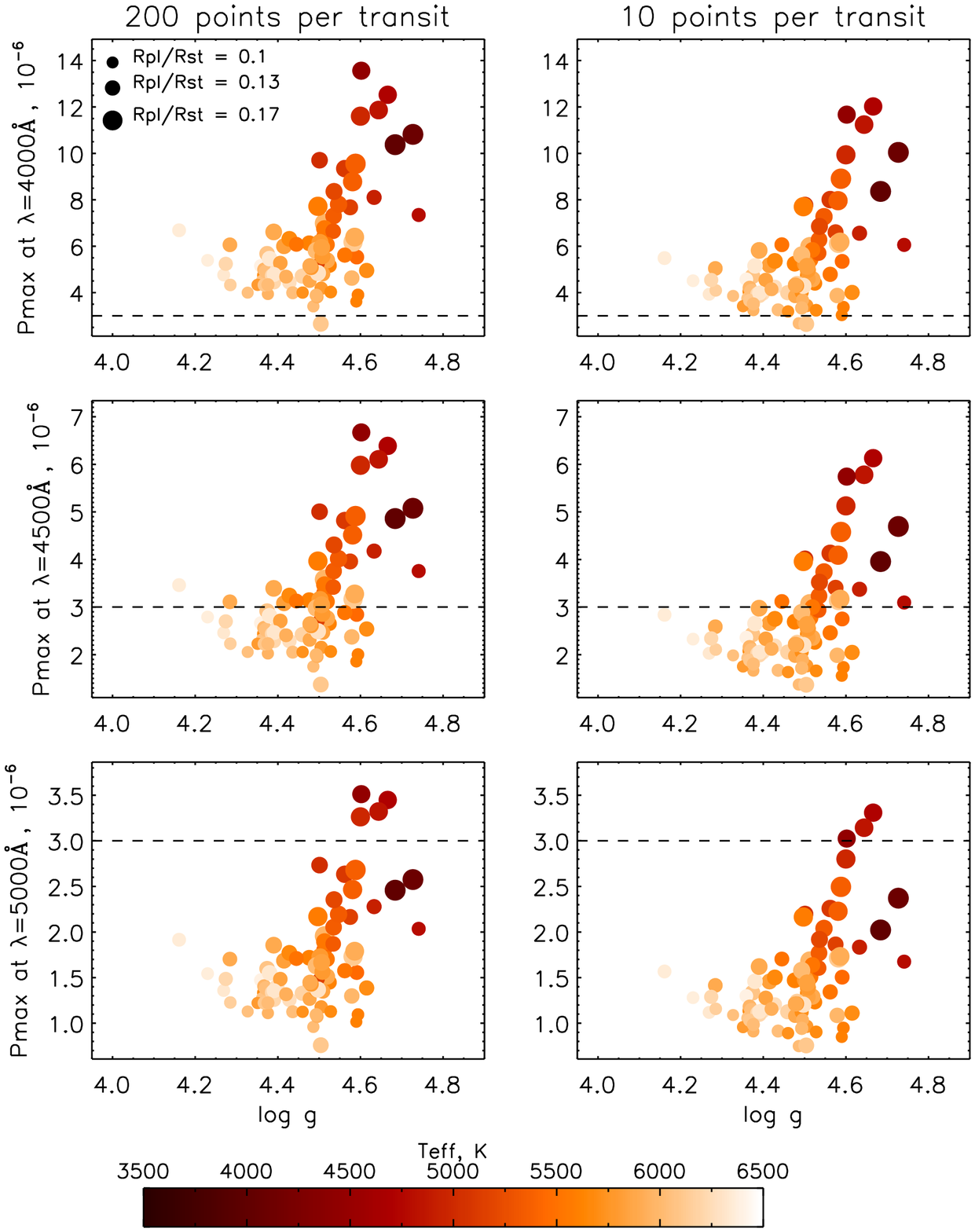}
\caption{Maximum polarization degree during transits for 88 exoplanetary systems as a function of
the surface gravity of the host stars. The gray scale refers to the effective 
temperature. The size of each circle designates the planet-to-star radii ratio
(in the range from 0.1 to 0.17). The top, middle and bottom plots show the maximum polarization degree
at the wavelengths of $4000\rm{\AA}$, $4500\rm{\AA}$ and $5000\rm{\AA}$, respectively. The left plots were
simulated
assuming 200 data points per transit and the right plots with 10 points per transit. The horizontal dashed line
marks the lower limit ($3\sigma$) of achievable polarization sensitivity.
(A colour version of this figure is
available in the online edition of the journal). 
\label{fig3}}
\end{figure}

In Figure \ref{fig3}, we plot the maximum polarization degree assuming 200 measurements per transit (left plots)
and 10 measurements per transit (right plots) versus the surface gravity (x-axis) for different effective temperatures
(colored scale) and planet-to-star radii ratios (scaled circle size) based on our sample of transiting exoplanets.
Three rows of plots correspond to different wavelengths of $ 4000\rm{\AA}$,  $4500\rm{\AA}$  and $5000\rm{\AA}$.
The horizontal dashed line marks the selected sensitivity limit of $P_{max} = 3\times 10^{-6}$.

\placefigure{fig3}

As seen from Fig. \ref{fig3}, if we assume 200 data points per transit, there are
more objects above the selected polarization sensitivity limit, as compared to 10 points. However, 
since the transit polarization is very small and the transit duration is usually
short, it is not yet possible to resolve 200 points per transit, while 10 points
per transit is feasible. Since linear polarization is larger in the blue
spectral range due to Rayleigh scattering and high 
temperature gradient in the stellar atmosphere \citep{harrington69, magalhaes86, kostogryz14}, the maximum
polarization degree during the transit is expected to be larger at shorter wavelengths
as well.

We chose several groups of exoplanets suitable for observations from the objects in 
Table \ref{tbl-1} with maximum polarization degree above the
adopted sensitivity limit at $4500 \rm{\AA}$.
In the first one we have the brightest transiting system, HD~189733, which
shows a maximum polarization degree of $3.43\times 10^{-6}$ for averaged 10 points 
per transit. Next, we have the group of objects with apparent magnitudes $m_V$ between 11.0 and 12.0:
HAT-P-20, Kepler-16 (AB), WASP-2, WASP-45, WASP-52 and WASP-80. Finally, the group with $m_V$  
between 12.0 and 13.0 includes:
Corot-2, HAT-P-18, HAT-P-19, Qatar-1, TrES-3, WASP-10, WASP-43, WASP-67 and WASP-98.
There are also four systems where the maximum polarization is larger than $10^{-5}$ at $\lambda  = 4000 \rm\AA$
which is very promising for detection.
Unfortunately, the transiting system HAT-P-54, which has the largest polarization,
is very faint, $m_V = 13.5$ and can be observed with a large telescope only.
In their recently submitted paper, \citet{wiktorowicz14} reported on failed attempt of detecting transit polarization
of HD~80606 because of the low sensitivity and systematics of their polarimetry. 
This system is not included in our sample, as it is not meeting our selection criteria.
However, we simulated the maximum polarization degree during its transit to be $1.5\times10^{-6}$
for the $4500~\rm \AA$ wavelength. Even accounting for all the telescope systematic effects, HD~80606
is not a suitable target for polarimetric transit observations with the current instrumentation.
Below we discuss in details three systems where detection of transit polarization is feasible and/or desirable.
    
\paragraph{WASP-43.}
It is a hot Jupiter transiting the K7V star every 0.81 days that was discovered by \citet{hellier11}.
With the large planet-to-star radii ratio of 0.159 and a cool host star ($T_{eff} = 4400K$),
WASP-43 is expected to have a large polarization degree of $1.1\times10^{-5}$ at $\lambda = 4000 \rm\AA$
during the transit and, therefore, it is 
one of the most promising objects for polarimetric observations. Due to its very short orbital period,
it would be possible to make many observations of the system transits and then combine them to get
a better sensitivity. However, having a quite faint host star with $m_V = 12.4$, a large
telescope is needed to observe this object with sufficient photon statistics.

\begin{figure}
\includegraphics[width=.6\linewidth,angle=-90]{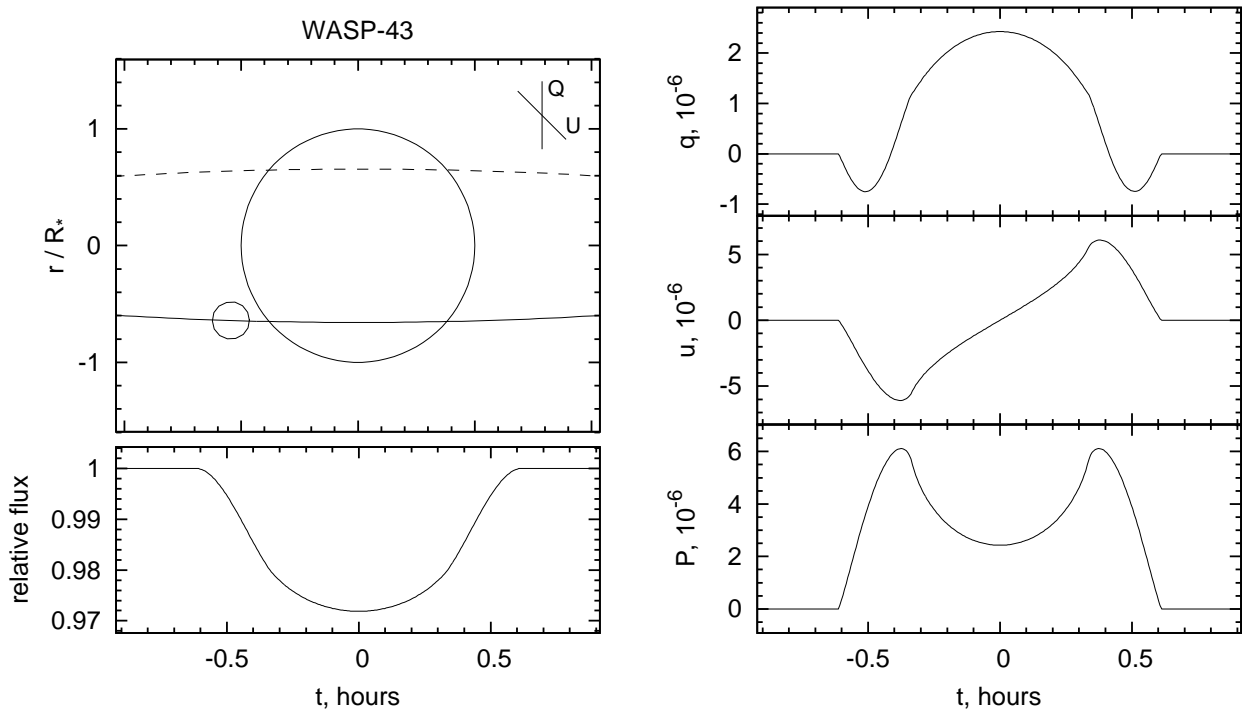}
\caption{The transiting system WASP-43 configuration and the simulated variations
of the flux and the polarization (q, u, and p) (assuming 200 points per transit) at $\lambda = 4500 \rm \AA$
and $\Omega = 90\deg$.
The star and the planet are shown in the top left panel as large and small circles,
respectively. The planet is orbiting from left to right in front of the star.
The dashed line indicates the rear part of the orbit. The positive sign directions of
Stokes q and u parameters are indicated in the top right corner on the same panel.
 \label{fig4}}
\end{figure}

\placefigure{fig4}

Figure \ref{fig4} presents the results of our simulations for WASP-43. The maxima of the polarization 
degree, listed in Table \ref{tbl-1} and plotted in Fig. \ref{fig3}, are observed at $\pm0.4$ hour
from the transit center. However, depending on $\Omega$ the maximum in q and u can also
occur near the transit centre.
  
\begin{figure}
\includegraphics[width=.6\linewidth,angle=-90]{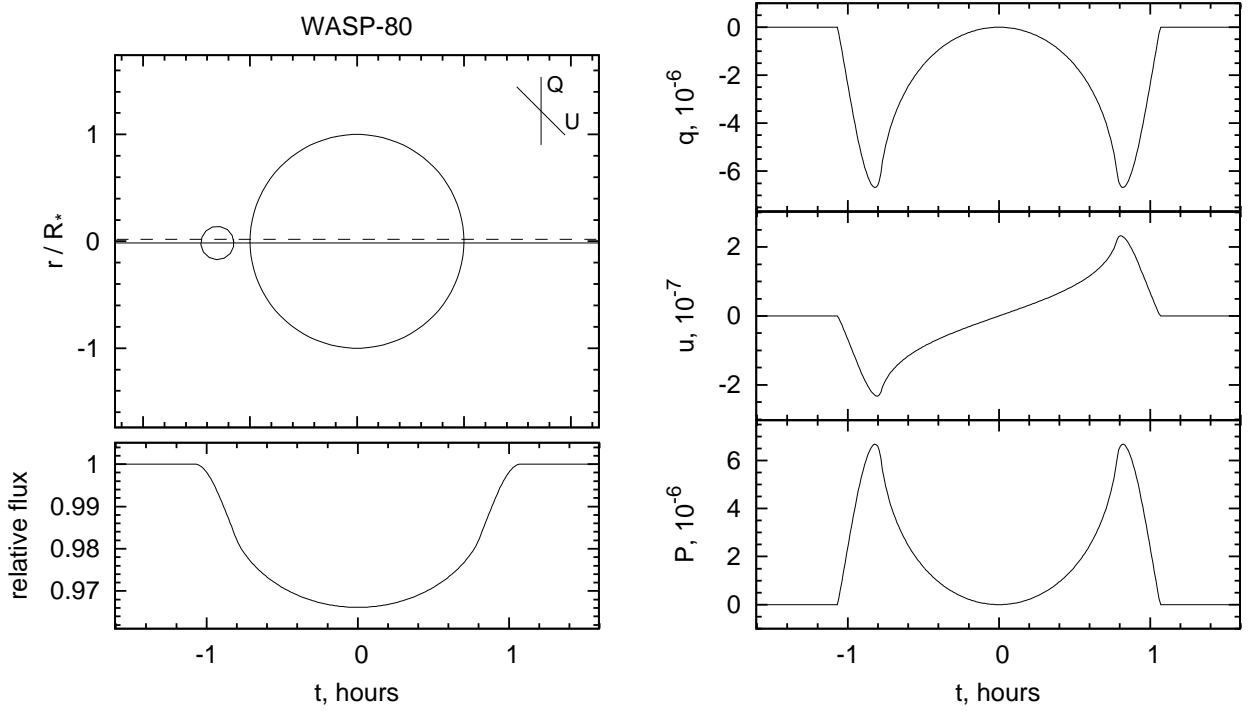}
\caption{ The same as in Figure \ref{fig4}  for WASP-80 \label{fig5}}.
\end{figure}

\begin{figure}
\includegraphics[width=.6\linewidth,angle=-90]{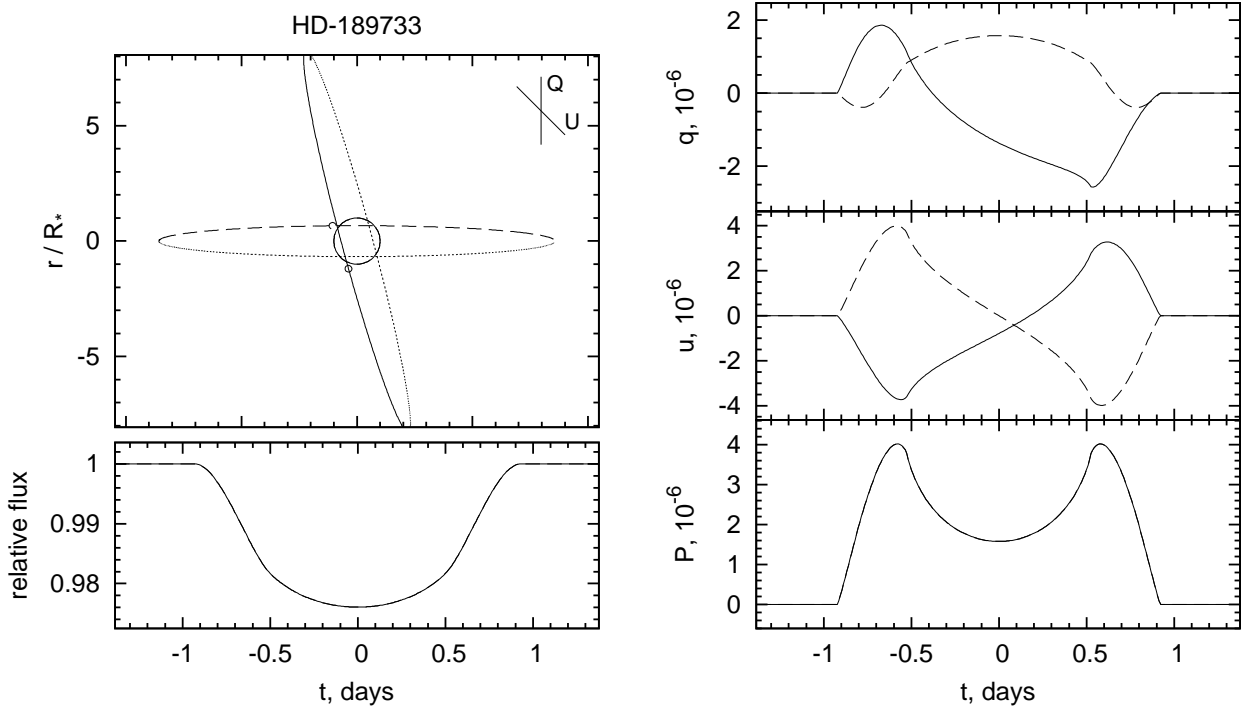}
\caption{The same as in Figure \ref{fig4} for HD~189733 for two different values of ascending node $\Omega$. 
In the upper left plot, the whole orbits of the planet are depicted. The solid lines correspond to the results with
$\Omega = 195\deg$ and and dashed lines are for  $\Omega = 90\deg$ (as assumed for all planets in this paper). \label{fig5.1}}
\end{figure}

\paragraph{Wasp-80.}
The discovery of the planet transiting the star WASP-80 was reported by \citet{triaud13}.
It has the effective temperature and the planet-to-star radii ratio that are similar to WASP-43. 
In Fig.\ref{fig5} we present the flux and the polarization degree variation during transit with 200 points for Wasp-80.
However, the difference between the simulated maximum polarization degree assuming 
10 and 200 data points per transit is more prominent for WASP-80.
This is explained by the different inclination angle: 
the larger the center-to-second or -third contact polarization variation, the sharper its maxima.
Since the averaging has a smoothing effect, it will be more pronounced for
orbits with larger $i$. In this respect, the planetary transits farther from a stellar center
 are more preferable for observations.

\placefigure{fig5}

\paragraph{HD 189733.}
This is currently one of the brightest ($m_V = 7.67$) star known to harbour
a transiting exoplanet \citep{bouchy05}. This, along with the short period
(2.2 days) and the large planet-to-star radii ratio ($R_p/R_\star=0.15$), makes it very
suitable for various kinds of observations including polarimetry \citep{berd08, berd11}.
From polarimetric observations 
of phase curves for HD~189733, \citet{berd11} obtained $\Omega\approx15\deg$ or $195\deg$ in the B band, 
because of $180 \deg$ ambiquity.  We take  $\Omega\approx195\deg$ and the same direction of the planet orbiting the star,
corresponding to inclination angle $ i = 94.49 \deg$ as in \citet{berd11}. 
In Fig. \ref{fig5.1} we show the variations of Stokes q and u for different $\Omega$ angles.
It was mentioned above that the flux and polarization degree do not depend on this orbital parameter,
but Stokes parameters do.

\placefigure{fig5.1}
 
Overall, we confirm the results from \citet{carciofi05} that most suitable for observations are the transiting
systems that have a large planet-to-star radii ratio, cool host star and moderate orbital
inclination. In addition, we present here the calculation of Stokes parameters q and u that are very sensitive 
to the longitude of ascending node of the planetary orbit. 

\section{Grazing and near-grazing transiting planets}

A grazing transit is defined as such that only a part of planetary disk transits the host star's disk.
This means that the second and the third contact points are missing from the
transit light curves, which makes it difficult to obtain accurate measurements of physical parameters
of the planetary system. 

\begin{figure}
\includegraphics[width=.6\linewidth,angle=-90]{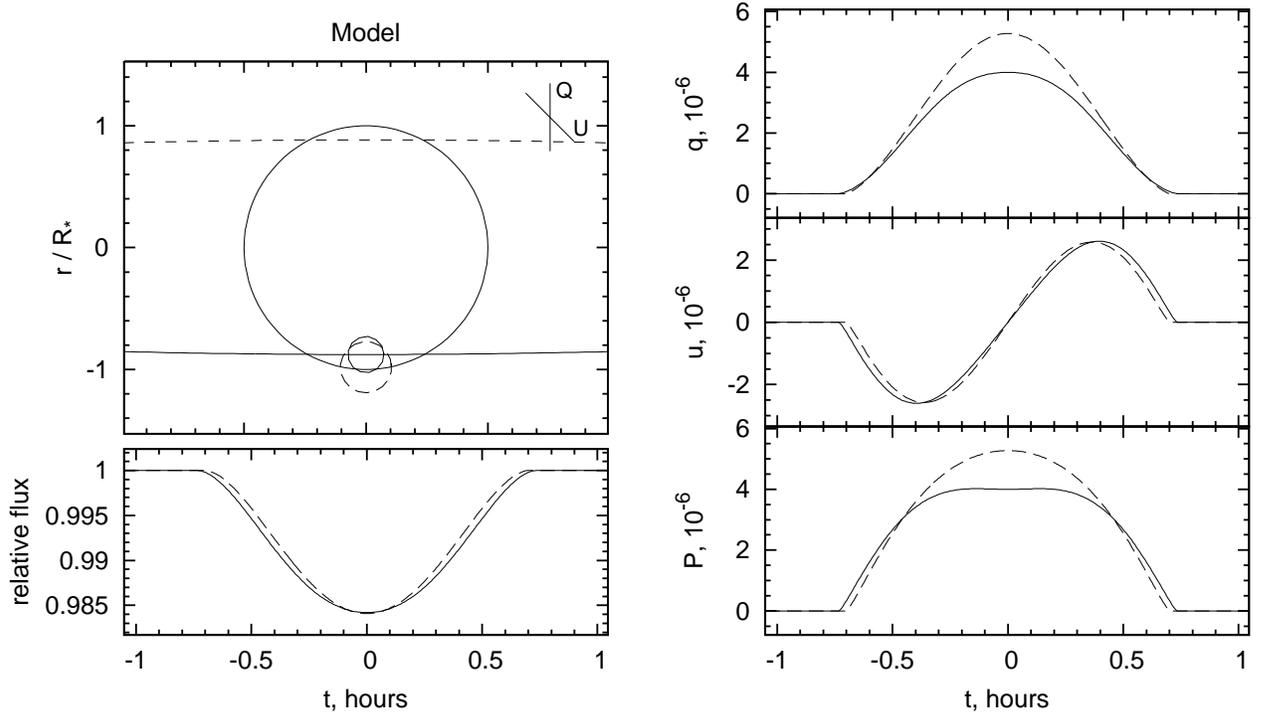}
\caption{ The same as in Figure \ref{fig4} for two models of grazing (dashed lines) 
and near-grazing (solid lines) transits. The radii for both planets are chosen in such a way that 
projected area, which blocks the stellar radiation is the same for the two cases.
\label{fig6}}
\end{figure}

Because stellar polarization is the largest at the limb, transit polarimetry can be particularly
useful for grazing or near-grazing transit observations.  
\citet{carciofi05} presented calculation for an example where the inclination angle for two planets
is the same while radii and projected area on the stellar disk, which blocks the radiation are different. 
In this case, they have large differences in flux and in polarization curves. Here 
we investigate how large is the difference in flux and polarization 
for grazing and near-grazing planetary transit when the planets block the same area on the stellar disk. 
In Fig. \ref{fig6},
we simulate two model systems with similar parameters and with the same projected area 
on the stellar disk, which is blocked by planet, while the inclination angles and
the planetary radii are different.  For a model star we take all parameters as in HD 189733. As HD 189733 is not grazing,
we decrease the inclination angle and planetary radius $r_1$ to simulate grazing transit. 
 The flux and the linear polarization variations for the first 
model of a near-grazing transiting system are shown with solid lines.
The larger model planet has $r_2 =  \sqrt{2}*r_1$ and inclination angle is chosen in such a way that
only half of the planet projected into stellar disk (dashed line in Fig. \ref{fig6}). 
 Evidently, it is hard to distinguish between these two
models based on the light curve only, as small difference in the wings can be
interpreted as variations of the stellar limb darkening. On the other hand, there
is a qualitative difference in the polarization degree variation curves: if the
whole planet crosses the star, a curve with two maxima is immediately established, while
the grazing transit reveals only one maximum in the center. This feature can be used
as good indicator to distinguish between the grazing and the near-grazing transits.

\placefigure{fig6}

In our sample in Table \ref{tbl-1}, we found two grazing planets WASP-34 and WASP-67, 
but only one of them (WASP-67) has the maximum polarization degree above the selected
polarimetric sensitivity limit. Additionally, we found several objects that have near-grazing
transits such as HAT-P-27, WASP-45 and TrES-3 (see online Fig. \ref{fig2.5}), which are also interesting from the polarimetric
point of view as in these cases the polarization signal should not change significantly
during the transit and one can average the data within one transit to get a better sensitivity.

\section{Planets transiting spotted stars}

Another effect which breaks the stellar disk symmetry and results in a non-zero linear
polarization is the stellar activity, considered here in terms of starspots.
There are several techniques by which starspots are characterized or characterize
the star \citep[e.g.,][]{berd05}. On the other hand, starspots are an additional
source of noise during transits when unaccounted for.
\citet{silva03} presented method to determine physical properties
of spots, such as size, intensity, position, and temperature using transiting
planets crossing over starspots. Later, \citet{silva-valio08} suggested that it would be possible to obtain 
the stellar rotation period from two consecutive transit observations, when the
planet passes in front of the same starspot. It was successfully simulated for the Sun
and applied for HD~209458, for which the rotation period was estimated of 11.4 days.
Spots can also provide the information about the stellar rotation axis spatial orientation,
i.e., stellar obliquity, usually expressed by a sky-projected spin-orbit misalignment
$\lambda$ \citep[e.g.,][]{sanchis-ojeda13a}. This technique also relies upon
detection of consecutive starspot-crossing events. 

A superposition of a planetary transit and starspots create a complex picture
of linear polarization signal variations. Importantly, starspots can be revealed with transit
polarimetry even in the absence of spot crossing events, thus giving a more complete account
of stellar activity and potentially complementing the aforementioned techniques.
Different spot locations contribute in different way to integral polarized light. 
Evidently, the polarization degree due to a near-limb spot is the
highest, and at other spot locations its contribution will be less significant.
Furthermore, starspot and planet polarization, when superimposed, can compensate
each other. According to a definition of Stokes q and u parameters, the maximum
compensation is observed when the angle between the planet and the spot as seen from the stellar
disk center is $\pm90\deg$. This effect is especially significant when there are
many spots distributed over the stellar surface, negating each other.
We have chosen two targets to demonstrate the starspots' influence on polarization during
planetary transits which are discussed in detail below.

\begin{figure}
\includegraphics[width=.6\linewidth,angle=-90]{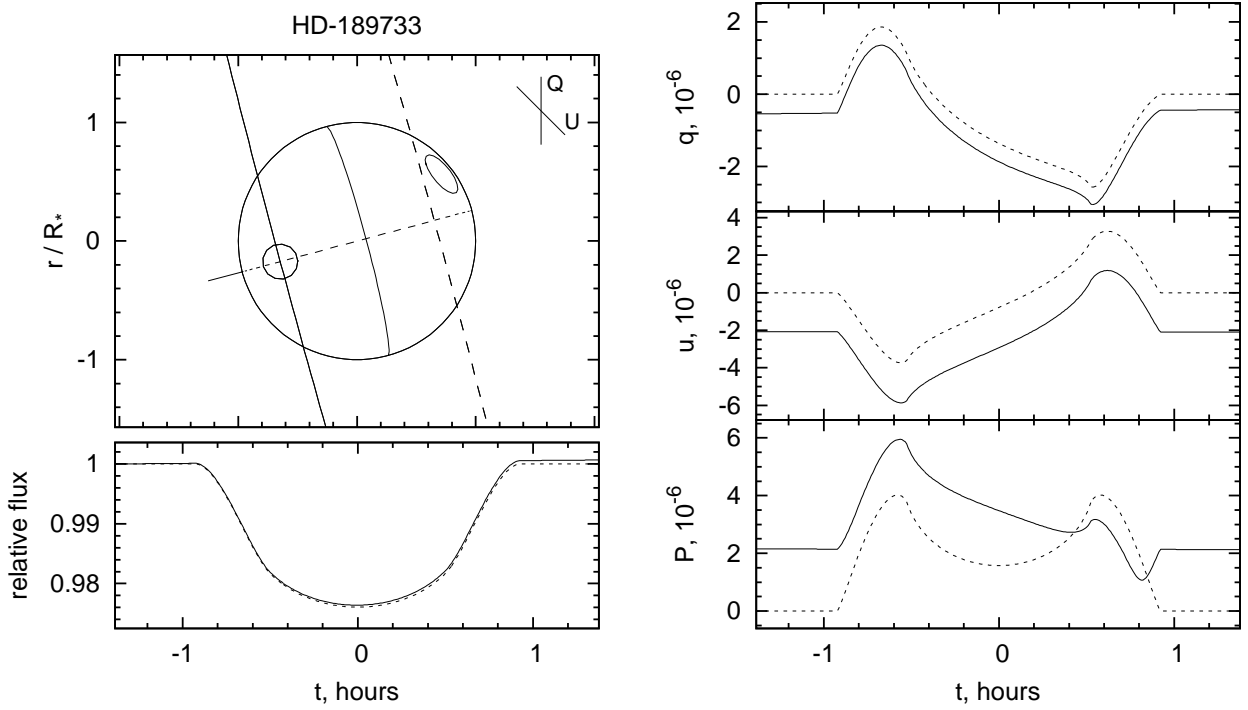}
\caption{The same as in Figure \ref{fig4} for the exoplanetary system HD~189733 ($\Omega = 195\deg$ \bfseries and $i = 94.49\deg$\mdseries)
with a spot added near the limb of the host star. The results are presented for the two transit cases: with the spot
(solid lines) and without spots (dotted lines). The planet is orbiting from bottom to top in
front of the star.\label{fig7}}
\end{figure}

\paragraph{HD 189733.}
It is an active star with spots covering up to $1\%$ of its surface \citep{boisse09, winn07}.
The star has a rotation period of 11.953 days, accurately estimated from optical photometry by 
\citet{henry08}. In our simulations we set different parameters for spots such as temperature
\citep[73\% of the star's effective temperature,][]{berd05}, size ($1\%$ of the stellar surface) and
positions (latitudes in the range of $30 - 70\deg$) to calculate the 
flux and polarization (q, u and p) variations during the planetary transit. One result
 when the entire $1~\%$ spotted area is placed near the limb of the star is shown in Fig. \ref{fig7} 
with solid lines. For comparison, the case of the transit
with no starspots is indicated by dotted lines. It is seen that the flux light curves for
these two cases are nearly identical showing only slight deviation at the end of the
transit. At the same time, the spot results in the noticeable vertical offset
of the polarization curves for q and u parameters. The total effect is even more
pronounced in the polarization degree, where, in addition to the offset, the shape of
the curve also changes having the second maximum lowered down. This demonstrates
the specifics of the polarimetric approach which deals with the azimuthal dependence of stellar 
limb polarization as compared to flux which is independent on azimuth.

From our simulation, we estimate a $1~\%$ starspot contribution to polarization for the extreme
limb case for HD~189733 to be $\sim2\times10^{-6}$ at $4500\rm{\AA}$. In another study,
\citet{berd11} selected the same extreme starspot parameters, but used different calculation
technique, as well as input limb-polarization variations. They obtained the
value of $3\times10^{-6}$ in the B filter, which is in a good agreement with our simulations.

\placefigure{fig7}

\paragraph{Kepler-63.} This giant planet orbiting a young Sun-like star
was discovered and characterized by \citet{sanchis-ojeda13b}. Its transits occur
every 9.43 days with apparent brightness variations caused by large starspots.
The host star rotates with a shorter period of 5.4 days. Based on the
Rossiter-McLaughlin effect and the analysis of starspot-crossing events, \citet{sanchis-ojeda13b}
determined that the star has the high obliquity of $\psi = 104\deg$. 
Note that Kepler-63 is not included in our sample (Table \ref{tbl-1}),
because its planet-to-star ratio of 0.06 is smaller than our selection criterium $R_p/R_\star=0.1$.
Nevertheless, because of the reported significant stellar activity, we reconstruct the
configuration of the system including a large polar spot
and calculate the flux and polarization parameter variations.

\begin{figure}
\includegraphics[width=.6\linewidth,angle=-90]{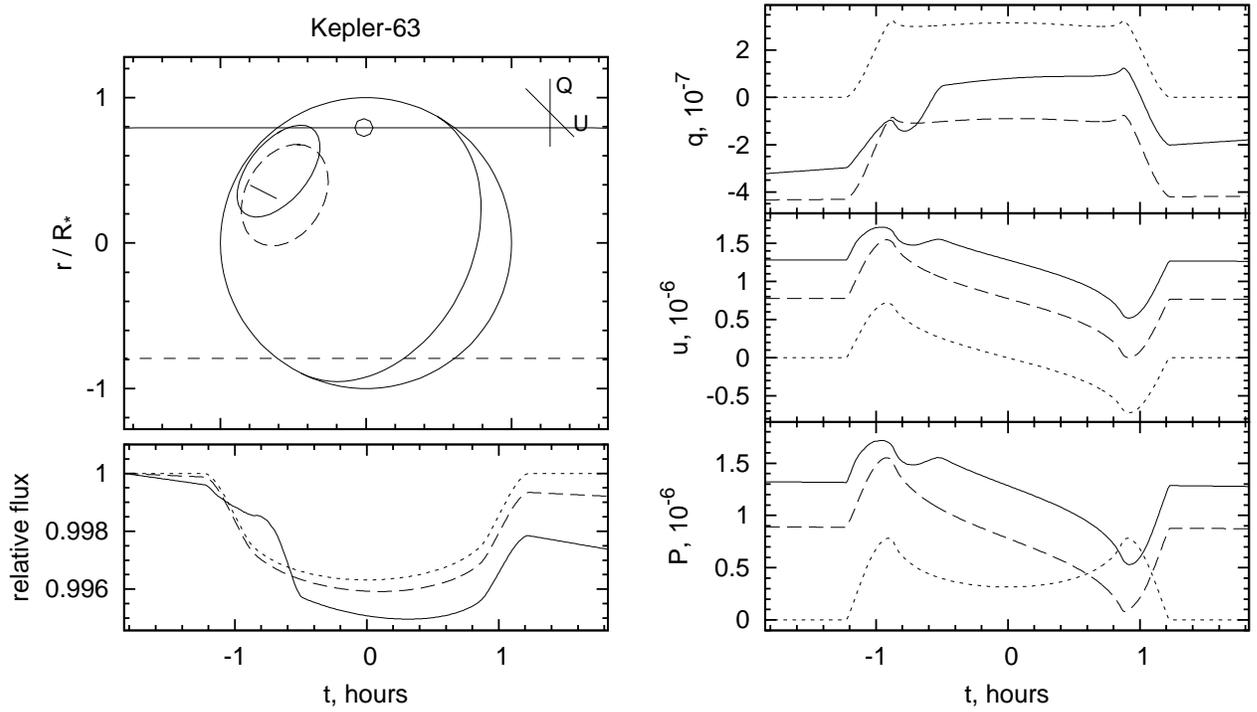}
\caption{The exoplanetary system Kepler-63 configuration and simulated variations of the flux and
the polarization for three different cases: the plain planetary transit (dashdotted lines),
the transit and a starspot outside planet path (dashed lines) and the transit with a spot-crossing event (solid lines).
\label{fig8}}
\end{figure}

In Figure \ref{fig8} we present the results of simulations for three different cases.
The first one, plain transit (dashdotted lines), is characterized by a shallow
light curve and small variations of the polarization parameters as compared to the
objects from our sample, because of the small size of the planet in Kepler-63.
Adding a large polar spot (spot area up to $5\%$ of the stellar surface) in the 
second case (dashed lines) outside the path of 
the planet results in a very similar light curve but
introduces significant vertical offsets of the q and u curves.
The polarization degree curve looks also different: it is shifted to large values
and has gained two oppositely-oriented maxima. As compared to the case in Fig. \ref{fig7}
for HD~189733, the larger starspot of Kepler-63 in the selected configuration
affects the polarization degree curve significantly more. Finally, in the third case
(shown with solid lines in Fig. \ref{fig8}) we set the starspot position so that
spot-crossing occurs. Since the spot is set now closer to the limb, we
see a further increase of the vertical offset of the polarization parameter curves.
At the same time the effect of the planet crossing the starspot which takes place between -0.8
and -0.5 hours is very small, and it is more noticeable in the light curve.
Note that detecting offsets of polarization curves requires the same polarimetric 
accuracy as the sensitivity (i.e., negligible systematic errors). However,
since the polarization degree is always non-negative the offset can be estimated from its shape.

\placefigure{fig8}


\section{Summary}

The numerical method presented here allows to simulate flux and linear
polarization variations in transiting exoplanetary systems. We considered two
cases of the stellar disk symmetry breaking: including planetary transits and
starspots, and estimated their combined effect.

We selected 88 objects from the Extrasolar Planets Encyclopedia and identified 
a sample of transit systems that should exhibit maximum polarization effect and,
thus, are most promising for future polarimetric observations. Depending on the
selected wavelength, there are more than a dozen found objects with simulated
maximum polarization degrees that should be already achievable with high
sensitivity polarimeters. We show that polarimetric observations can
provide valuable parameters of the planetary systems, some of
which, like orbit spatial orientation, are undetectable by transit photometry. In addition,
transit polarimetry can characterize the host star as well, including 
center-to-limb variation of the polarization that can be used for testing
stellar models.

Polarimetry allows to distinguish between the grazing and near-grazing planetary transit.
Our simulations showed that there is a qualitative difference in the polarization degree
curves for these two cases, while photometric curves are nearly the same in shape.
Moreover, for a near-grazing
exoplanetary transit, the polarization degree will be almost constant during the transit.
This allows to average the polarimetric signal from almost entire transit for a better sensitivity.
In our sample, we found one grazing system WASP-67, which has maximum polarization degree
above the chosen sensitivity limit of $3\times10^{-6}$.

Based on the existing studies of stellar activity, we analysed the system
configurations of HD~189733 and Kepler-63 and simulated the flux and linear polarization
variations with starspots, including spot-crossing events. Our results indicated
that starspots can contribute significantly to the polarization, by causing vertical offsets in 
the curves and, in particular, changing the shape of
the polarization degree variation even if the spot is not crossed by the planet, 
while remaining hardly detectable in the relative
flux light curves. The important finding is that polarization parameters are sensitive to spot
sizes, positions and reveal the stellar rotation period equally with and
without spot-crossing effect.

\begin{acknowledgements}
      This work was supported by the European Research Council Advanced Grant HotMol (ERC-2011-AdG 291659).
      We thank our referee Dr. Antonio (Mario) Magalh\~aes for useful corrections and suggestions that have improved
      our paper.
      
\end{acknowledgements}

\end{document}